\documentclass[runningheads]{svmult}

\usepackage{graphicx}  
\usepackage{physprbb}  

\def\gsim{\mathrel{\raise.3ex\hbox{$>$\kern-.75em\lower1ex\hbox{$\sim$}}}}

\begin{document}

\font\fortssbx=cmssbx10 scaled \magstep1

\title*{\vspace*{-6ex}\hbox to \hsize{{\fortssbx University of Wisconsin - Madison}
\hfill\vtop{\normalsize
\hbox{\bf MADPH-02-1277}
\hbox{May 2002}
\hbox{\hfil}}}
High-energy Neutrinos from Cosmic Rays\vspace{-3ex}\footnote{Presented at the  ESO-CERN-ESA Symposium on  Astronomy, Cosmology and Fundamental Physics,  Garching, Germany, March 4--7, 2002}}
\toctitle{High-energy Neutrinos from Cosmic Rays}
\titlerunning{High-energy Neutrinos from Cosmic Rays}

\author{\vspace{-2ex}Francis Halzen}

\authorrunning{Francis Halzen}

\institute{Department of Physics, University of Wisconsin,
Madison, WI 53706}

\maketitle
\vspace*{-8ex}

\begin{abstract}

We introduce neutrino astronomy from the observational fact that Nature
accelerates protons and photons to energies in excess of $10^{20}$ and
$10^{13}$\,eV, respectively. Although the discovery of cosmic rays dates
back close to a century, we do not know how and where they are
accelerated. We review the facts as well as the speculations about the
sources. Among these gamma ray bursts and active galaxies represent
well-motivated speculations because these are also the sources of the
highest energy gamma rays, with emission observed up to 20\,TeV,
possibly higher.

We discuss why cosmic accelerators are also expected to be cosmic beam
dumps producing high-energy neutrino beams associated with the highest
energy cosmic rays. Cosmic ray sources may produce neutrinos from MeV to
EeV energy by a variety of mechanisms. The important conclusion is that,
independently of the specific blueprint of the source, it takes a
kilometer-scale neutrino observatory to detect the neutrino beam
associated with the highest energy cosmic rays and gamma rays. The
technology for commissioning such instruments exists.

\end{abstract}

\section{The Highest Energy Particles: Cosmic Rays, Photons and
Neutrinos}

\subsection{The New Astronomy}

While conventional astronomy spans 60 octaves in photon frequency, from
$10^4$\,cm radio-waves to $10^{-14}$\,cm  gamma rays of GeV energy,
successful efforts are underway to probe the Universe at yet smaller
wavelengths  and larger photon energies; see Fig.\,1. Gamma rays,
gravitational waves, neutrinos and very high-energy protons are explored
as astronomical messengers. As exemplified time and again, the
development of novel ways of looking into space invariably results in
the discovery of unanticipated phenomena. As is the case with new
accelerators,
observing only the predicted will be slightly disappointing.

\begin{figure}[t!]
\centering\leavevmode
\includegraphics[height=3in]{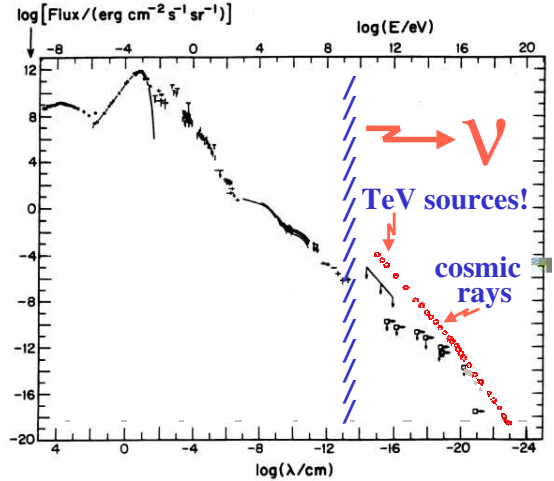}
\caption{The diffuse flux of photons in the Universe, from radio waves to
GeV-photons. Above tens of GeV, only limits are reported although
individual sources emitting TeV gamma rays have been identified. Above
GeV
energy, cosmic rays dominate the spectrum.}
\label{one}
\end{figure}

Why pursue high-energy astronomy with neutrinos or protons despite the
considerable instrumental challenges? A mundane
reason is that the
Universe is not transparent to photons of TeV energy and above
(in ascending factors of $10^3$, units are: GeV/TeV/PeV/EeV/ZeV ). For
instance, a PeV energy photon cannot deliver information from a
source
at the edge of our own galaxy because it will annihilate into an
electron pair in an encounter with a 2.7 Kelvin microwave
photon before reaching our telescope. Only neutrinos can reach us
without attenuation from the edge of the Universe at all energies.

At EeV energies, proton astronomy may be possible. Above 50\,EeV the
arrival directions of electrically charged cosmic rays are
no longer scrambled by the ambient magnetic field of our own
galaxy.    
They point back to their sources with an accuracy determined by their
gyroradius in the intergalactic magnetic field $B$:
\begin{equation}
{\theta\over0.1^\circ} \cong { \left( d\over 1{\rm\ Mpc} \right)
\left( B\over 10^{-9}{\rm\,G} \right) \over \left( E\over
3\times10^{20}\rm\, eV\right) }\,,
\end{equation}
where $d$ is the distance to the source. Speculations on the strength of
the inter-galactic magnetic field
range from $10^{-7}$ to $10^{-12}$~Gauss in the local cluster. For a
distance of 100~Mpc,
the resolution may therefore be anywhere from sub-degree to
nonexistent. Proton astronomy should
be possible at the very highest energies; it may also provide indirect
information on
intergalactic magnetic fields. Determining the strength of intergalactic
magnetic fields by conventional astronomical means has been challenging.

\subsection{The Highest Energy Cosmic Rays: Facts}

In October 1991, the Fly's Eye cosmic ray detector recorded an event
of energy $3.0\pm^{0.36}_{0.54}\times 10^{20}$\,eV \cite{flyes}.
This event, together with an event recorded by the Yakutsk air shower
array in May 1989 \cite{yakutsk}, of estimated energy $\sim$
$2\times10^{20}$\,eV, constituted at the time the two highest
energy cosmic rays ever recorded. Their energy corresponds to a center of
mass energy of the order of 700~TeV or $\sim 50$ Joules, almost 50
times the energy of the Large Hadron Collider (LHC). In fact, all
active 
experiments \cite{web} have detected
cosmic rays in the vicinity of 100~EeV since their initial discovery by
the
Haverah Park air shower array \cite{WatsonZas}. The AGASA air shower
array in Japan\cite{agasa} has now accumulated an impressive 10
events with energy in excess of $10^{20}$\,eV \cite{ICRC}.

\begin{figure}[t!]
\centering\leavevmode
\includegraphics[height=3in]{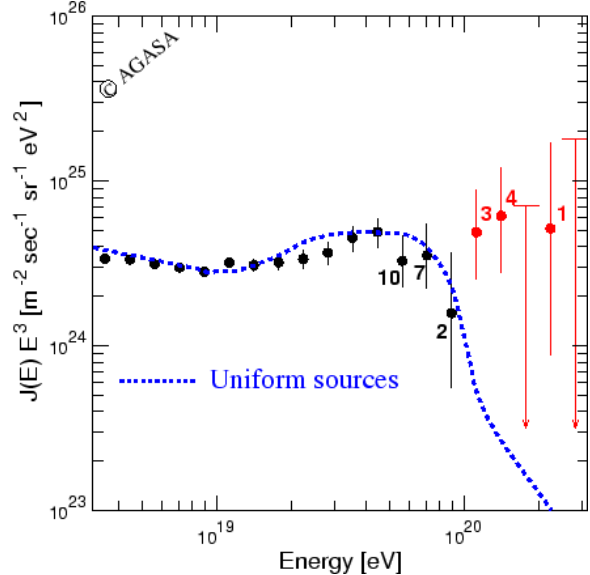}
\caption{The cosmic ray spectrum peaks in the vicinity of 1\,GeV and has
features near $10^{15}$ and $10^{19}$\,eV referred to as the ``knee" and
``ankle" in the spectrum, respectively. Shown is the flux of the highest
energy cosmic rays near and beyond the ankle measured by the AGASA
experiment.  Note that the flux is multiplied by $E^3$.}
\label{two}
\end{figure}

\begin{figure}[t!]
\centering\leavevmode
\includegraphics[height=3in]{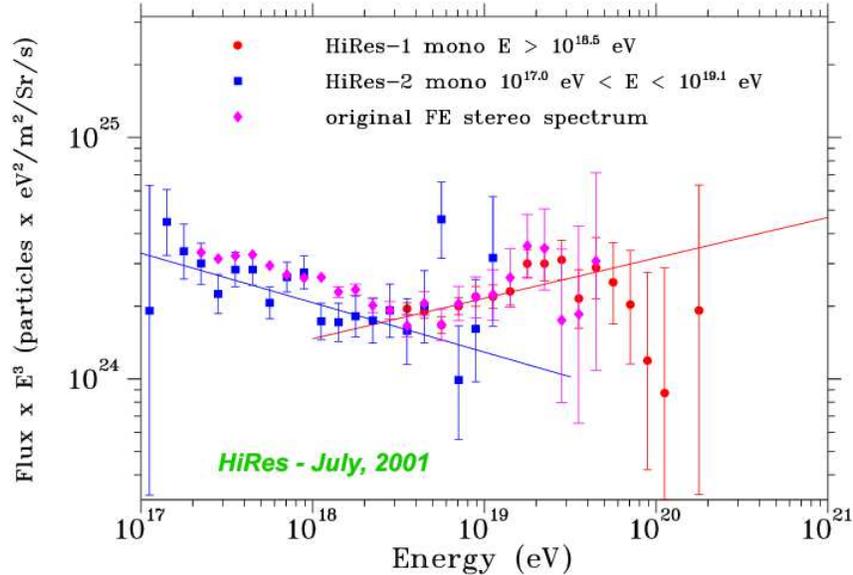}
\caption{As in Fig.\,2, but as measured by the HiRes experiment.}
\label{three}
\end{figure}

The accuracy of the energy resolution of these experiments is a critical
issue. With a particle flux of order 1 event per km$^2$ per
century, these events are studied by using the earth's
atmosphere as a particle detector. The experimental signature of an
extremely high-energy cosmic particle is
a shower initiated by the particle. The primary particle creates an
electromagnetic and
hadronic cascade.  The electromagnetic shower grows to a shower maximum,
and is subsequently absorbed by the
atmosphere. The shower can be observed by: i)
sampling the electromagnetic and hadronic components when they reach
the ground
with an array of particle detectors such as scintillators, ii)
detecting the fluorescent
light emitted by atmospheric nitrogen excited by the passage of the
shower particles, iii) detecting the Cerenkov light emitted by the
large number
of particles at shower maximum, and iv)~detecting muons and neutrinos
underground.

The bottom line on energy measurement is that, at this time, several
experiments using the first two techniques agree on the energy of
EeV-showers within a typical resolution of 25\%. Additionally, there is a
systematic error of order 10\% associated with the modeling of the
showers. All techniques are indeed subject to the ambiguity of particle
simulations that involve physics beyond the LHC. If the final outcome
turns out to be an erroneous inference of the energy of the shower
because
of new physics associated with particle interactions at the
$\Lambda_{\mathrm{QCD}}$ scale, we will be happy to contemplate this
discovery
instead.

The premier experiments, HiRes and AGASA, agree that cosmic rays with
energy in excess of 10\,EeV are not galactic in origin and that
their spectrum extends beyond 100\,EeV. They disagree on almost
everything else. The AGASA experiment claims evidence that the highest
energy cosmic rays come
from point sources, and
that they are mostly heavy nuclei. The HiRes data does not support
this. Because of low statistics, interpreting the measured fluxes as
a function of energy is like reading tea leaves; one cannot help however
reading different messages in the spectra (see Fig.\,2 and Fig.\,3).

\subsection{The Highest Energy Cosmic Rays: Fancy}

\subsubsection*{Acceleration to $>100$ EeV?}

It is sensible to assume that, in order to accelerate a proton to
energy $E$ in a magnetic field $B$, the size $R$ of the accelerator
must be larger than the gyroradius of the particle:
\begin{equation}
R > R_{\rm gyro} = {E\over B}\,.
\end{equation}
That is, the accelerating magnetic field must contain the particle
orbit. This condition yields a maximum energy
\begin{equation}
E \sim \gamma BR
\end{equation}
by dimensional analysis and nothing more. The $\gamma$-factor has
been included to allow for the possibility that we may not be at rest
in the frame of the cosmic accelerator.  The result would be the
observation of boosted particle energies.
Theorists' imagination regarding the accelerators has been limited to
dense regions where exceptional gravitational forces
create relativistic particle flows: the dense cores of exploding
stars, inflows on supermassive black holes at the centers of active
galaxies, annihilating black holes or neutron stars. All
speculations involve collapsed objects and we can therefore replace $R$
by the Schwartzschild radius
\begin{equation}
R \sim GM/c^2
\end{equation}
to obtain
\begin{equation}
E \propto \gamma BM \,.
\end{equation}
Given the microgauss magnetic field of our galaxy, no structures are
large or massive enough to reach the energies of the highest energy
cosmic rays. Dimensional analysis therefore limits their sources to
extragalactic objects; a few common speculations are listed in
Table\,1.

Nearby active galactic nuclei, distant by $\sim100$~Mpc and
powered by a billion solar mass black holes, are candidates.
With kilogauss fields, we reach 100\,EeV. The jets (blazars) emitted by
the
central black hole could reach similar energies in accelerating
substructures (blobs) boosted in our direction by Lorentz factors of 10
or
possibly higher. The neutron star or black hole remnant of a
collapsing
supermassive star could support magnetic fields of $10^{12}$\,Gauss,
possibly larger. Highly relativistic shocks with $\gamma > 10^2$
emanating
from the
collapsed black hole could be the origin of gamma ray bursts and,
possibly, the source of the highest energy cosmic rays.

\begin{table}[t]
\caption{Requirements to generate the highest energy cosmic rays in
astrophysical sources.}
\centering\leavevmode
\begin{tabular}{|llll|}
\hline
\multicolumn{4}{|c|}{Conditions with $E \sim 10\rm\ EeV$}\\
\hline
$\bullet$\ Quasars& $\gamma\cong 1$& $B\cong 10^3$ G& $M\cong 10^9
M_{\rm sun}$\\
$\bullet$\ Blazars& $\gamma\gsim 10$& $B\cong 10^3$ G& $M\cong 10^9
M_{\rm sun}$\\
$\bullet$\ \parbox[t]{1.0in}{Neutron Stars\\ Black Holes\\ $\vdots$}&
$\gamma\cong 1$& $B\cong 10^{12}$ G& $M\cong M_{\rm sun}$\\
$\bullet$\ GRB& $\gamma\gsim 10^2$& $B\cong 10^{12}$ G& $M\cong M_{\rm
sun}$\\
\hline
\end{tabular}
\end{table}

The above speculations are reinforced by the fact that the sources
listed are also the sources of the highest energy gamma rays
observed. At this point, however, a
reality check is in order. The above
dimensional analysis applies to the Fermilab accelerator: 10 kilogauss
fields over several kilometers corresponds to 1\,TeV. The argument holds
because, with optimized design and perfect alignment of magnets, the
accelerator
reaches efficiencies matching the dimensional limit. It is highly
questionable that nature can achieve this feat. Theorists can
imagine acceleration in shocks with an efficiency of perhaps 10\%.

The astrophysics of accelerating particles to Joule energies is so
daunting that many believe that
cosmic rays are not the beams of cosmic accelerators but the decay
products of remnants from the early Universe, such as topological
defects associated with a Grand Unified Theory (GUT) phase transition.

\subsubsection*{Are Cosmic Rays Really Protons: the GZK Cutoff?}

All experimental signatures agree on the particle nature of the
cosmic rays --- they look like protons or, possibly, nuclei. We
mentioned at the beginning of this article that the Universe is
opaque to photons with energy in excess of tens of TeV because they
annihilate into electron pairs in interactions with the cosmic microwave
background.
Protons also interact with background
light, predominantly by photoproduction of the $\Delta$-resonance,
i.e.\ $p + \gamma_{CMB} \rightarrow \Delta \rightarrow \pi + p$ above
a threshold energy $E_p$ of about 50\,EeV given by:
\begin{equation}
2E_p\epsilon > \left(m_\Delta^2 - m_p^2\right) \,.
\label{eq:threshold}
\end{equation}
The major source of proton energy loss is photoproduction of pions on
a target of cosmic microwave photons of energy $\epsilon$. The
Universe is, therefore, also opaque to the highest energy cosmic rays,
with an absorption length of
\begin{eqnarray}
\lambda_{\gamma p} &=& (n_{\rm CMB} \, \sigma_{p+\gamma_{\rm
CMB}})^{-1}\\
&\cong& 10\rm Mpc,
\end{eqnarray}
when their energy exceeds 50\,EeV. This
so-called GZK cutoff establishes a universal upper limit on
the energy of the cosmic rays. The cutoff is robust,
depending only on two known numbers: $n_{\mathrm{CMB}} = 400\rm\,cm^{-3}$
and
$\sigma_{p+\gamma_{\mathrm{CMB}}} = 10^{-28}\rm\,cm^2$
   \cite{gzk1,gzk2,gzk3,gzk4}.

Cosmic rays do reach us with energies exceeding 100\,EeV. This presents
us with three options: i) the protons are
accelerated in nearby sources, ii)~they do reach us from distant sources
which accelerate them to even higher energies than we observe, thus
exacerbating the acceleration problem, or iii) the highest energy cosmic
rays are not protons.

The first possibility raises the challenge of finding an appropriate
accelerator by confining these already unimaginable sources to our
local galactic cluster. It is not impossible that all cosmic rays are
produced by the active galaxy M87, or by a nearby gamma ray burst
which exploded a few hundred years ago.

Stecker \cite{stecker2} has speculated that the highest energy cosmic
rays are Fe nuclei with a delayed GZK cutoff. The details are
complicated but the relevant quantity in the problem is $\gamma=E/AM$,
where A is the atomic number and M the nucleon mass. For a fixed
observed energy, the smallest boost towards GZK threshold is associated
with the largest atomic mass, i.e.~Fe.

\begin{figure}[t!]
\centering\leavevmode
\includegraphics[height=3in]{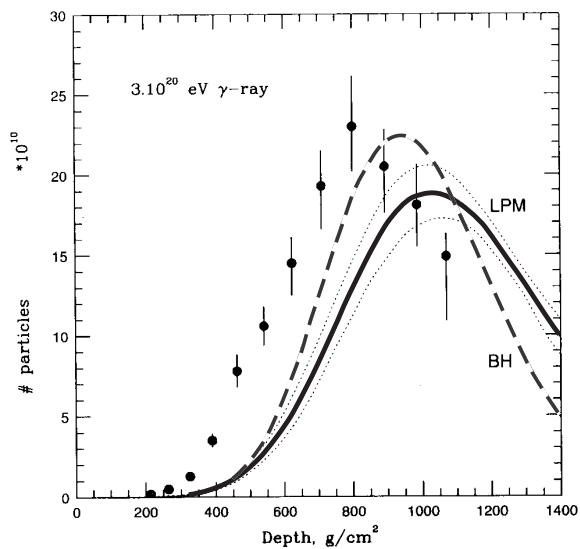}
\caption{The composite atmospheric shower profile of a $3\times
10^{20}$\,eV gamma ray shower calculated with Landau-Pomeranchuk-Migdal
(dashed) and
Bethe-Heitler (solid)
electromagnetic cross sections. The central line shows the average shower
profile and the upper and lower lines show 1~$\sigma$ deviations --- not
visible for the BH case, where lines overlap. The experimental shower
profile
is shown with the data points. It does not fit the profile of a photon
shower.}
\label{four}
\end{figure}

\subsubsection*{Could Cosmic Rays be Photons or Neutrinos?}

Above question naturally emerges in the context of models where the
highest energy cosmic rays are the decay products of remnants or
topological structures created in the early universe with typical energy
scale of order $10^{24}$\,eV. In these scenarios the highest energy
cosmic rays are predominantly photons. A topological defect will suffer
a chain decay into Grand Unified Theory (GUT) particles X and Y, that
subsequently
decay to familiar weak bosons, leptons and quark or gluon jets. Cosmic
rays are, therefore, predominately the fragmentation products of these
jets. We know from accelerator studies that, among the fragmentation
products of jets, neutral pions (decaying into photons) dominate, in
number, protons by close to two orders of magnitude. Therefore, if the
decay of topological defects is the source of the highest energy cosmic
rays, they must be photons. This is a problem because there is
compelling evidence that the highest energy cosmic rays are not photons:

\begin{enumerate}
\item The highest energy event observed by Fly's Eye is not likely to
be a photon \cite{vazquez}.  A photon of 300\,EeV will interact with the
magnetic field of the earth far above the atmosphere and disintegrate
into lower energy cascades --- roughly ten at this particular energy.
The detector subsequently collects light produced by the fluorescence of
atmospheric nitrogen along the path of the high-energy showers
traversing the atmosphere. The atmospheric shower profile of a 300\,EeV
photon after fragmentation in the earthÕs magnetic field, is shown in
Fig.\,4. It disagrees with the data. The observed shower
profile does fit that of a primary proton, or,
possibly, that of a nucleus. The shower profile information is
sufficient,
however, to conclude that the
event is unlikely to be of photon origin.

\item The same conclusion is
reached for the Yakutsk event that is characterized by a huge number
of secondary muons, inconsistent with a pure electromagnetic cascade
initiated by a gamma ray.

\item The AGASA collaboration claims evidence
for ``point" sources above 10\,EeV. The arrival directions are
however smeared out in a way consistent with primaries deflected by
the galactic magnetic field. Again, this indicates charged primaries
and excludes photons.

\item Finally, a recent reanalysis of the Haverah Park disfavors photon
origin of the primaries \cite{WatsonZas}.

\end{enumerate}

Neutrino primaries are definitely ruled out. Standard model neutrino
physics is understood, even for EeV energy. The average $x$ of
the parton mediating the neutrino interaction is of
order $x \sim \sqrt{M_W^2/s} \sim 10^{-6}$ so that the perturbative
result for the neutrino-nucleus cross section is calculable from measured
HERA structure functions. Even at 100\,EeV a
reliable value of the cross section can be obtained based on QCD-inspired
extrapolations of the structure function. The neutrino cross section is
known to better than an order of magnitude. It falls 5 orders of
magnitude
short of the strong cross sections required to make a neutrino interact
in
the upper atmosphere to create an air shower.

Could EeV neutrinos be strongly interacting because of new physics?
In theories with TeV-scale gravity, one can imagine that graviton
exchange dominates all interactions and thus erases the difference
between
quarks and
neutrinos at the energies under consideration. The actual models
performing this feat require a fast turn-on of the cross
section with energy that violates S-wave unitarity
\cite{han1,han2,han3,han4,han5,han6,han7,han8,han9}.

We have exhausted the possibilities.  Neutrons, muons and other
candidate primaries one may think of are unstable. EeV neutrons
barely live long enough to reach us from sources at the edge of our
galaxy.

\section{A Three Prong Assault on the Cosmic Ray Puzzle}

We conclude that, where the highest energy cosmic rays are concerned,
both
the
accelerator mechanism and the particle physics are enigmatic. The mystery
has inspired a worldwide effort to tackle the
problem with novel experimentation including air shower arrays covering
an area of several times $10^3$ square kilometers\cite{watson} and
arrays of multiple air Cerenkov telescopes\cite{volk}. We here discuss
kilometer-scale neutrino observatories. While these
have additional missions such as the search for dark matter\cite{tao},
their observations are likely to have an impact on cosmic ray physics.

\begin{figure}[t!]
\centering\leavevmode
\includegraphics[height=3in]{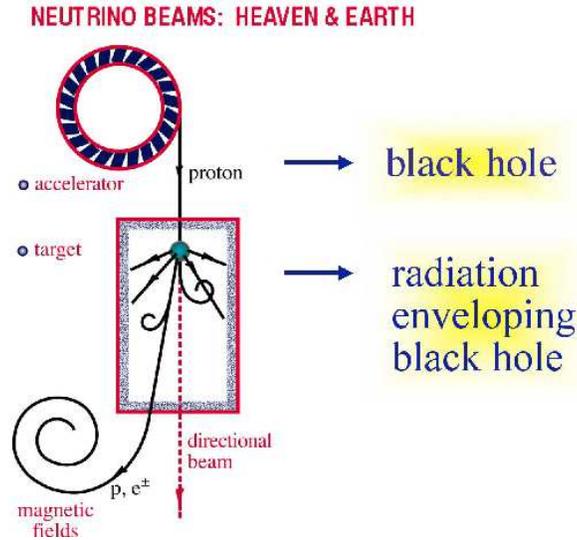}
\caption{Diagram of cosmic accelerator and beam dump.  See text for
discussion.}
\label{five}
\end{figure}

Why we anticipate that secondary photons and neutrinos are associated
with the highest energy cosmic rays is sketched in Fig.\,5. The cartoon
draws our attention to the fact that cosmic accelerators are also cosmic
beam dumps that produce secondary
photon and neutrino beams.
Accelerating particles to TeV energy and above requires relativistic,
massive bulk flows. These are likely to originate from the
exceptional gravitational forces associated with black holes or
neutron stars. Accelerated particles therefore pass through intense
radiation fields or dense
clouds of gas surrounding the black hole leading to the production of
secondary pions. These subsequently decay into photons and neutrinos
that accompany the primary cosmic ray beam. Example of beam dumps
include the external photon clouds or the UV radiation field that
surrounds the central black hole of active galaxies, or the matter
falling into the collapsed core of a dying supermassive star producing a
gamma ray burst.
The target material, whether a gas of particles or of photons,
is likely to be sufficiently tenuous for the primary proton beam
and the secondary photon beam to be only partially attenuated. However,
shrouded
sources from which only neutrinos can emerge, as in
terrestrial beam dumps at CERN and Fermilab, are also a possibility.

How many neutrinos are produced in association with the cosmic ray beam?
The answer to this question, among many
others\cite{snowmass1,snowmass2},  provides the rational for building
kilometer-scale neutrino detectors.

Let's first consider the question for the accelerator beam producing
neutrino
beams at an accelerator laboratory. Here the target absorbs all parent
protons as well as the muons, electrons and gamma rays (from $\pi^0
\rightarrow \gamma + \gamma$)
produced. A pure neutrino beam exits the dump. If nature constructed
such a ``hidden source" in the heavens,
conventional astronomy has not revealed it. It cannot be the source of
the
cosmic rays, however, for which the dump must be partially transparent to
protons.

In the other extreme, the accelerated proton interacts once, thus
producing the observed high-energy gamma rays \cite{cangoroo}.  It
subsequently escapes
the dump. We refer to this as a transparent source without absorption.
Particle physics
directly relates the number of
neutrinos to the number of observed cosmic rays and gamma
rays\cite{halzenzas}. Every observed cosmic
ray
interacts
once, and only once, to produce a neutrino beam determined only by
particle physics. The neutrino flux for such a transparent cosmic ray
source is referred to as the
Waxman-Bahcall flux \cite{wb1,wb2,R1,R2} and is shown as the horizontal
lines labeled ``W\&B" in
Fig.\,6. The calculation is valid for $E\simeq100$\,PeV. If the flux is
evaluated at both lower and higher cosmic ray energies, however, larger
values are found. This is shown as the non-flat line labeled
``transparent" in Fig.\,6. On
the lower side, the neutrino flux is higher because it is normalized to a
larger cosmic ray flux. On the higher side, there are more cosmic
rays in the dump to produce neutrinos because the observed flux at Earth
has been
reduced by absorption on microwave photons, the GZK-effect. The increased
values of the neutrino flux are also shown in Fig.\,6. The gamma ray
flux of $\pi^0$ origin associated with a transparent source is
qualitatively at the level of observed flux of non-thermal TeV gamma
rays from individual sources\cite{halzenzas}.

Nothing prevents us, however, from imagining heavenly beam dumps with
target densities somewhere between those of hidden and transparent
sources. When
increasing the target photon density, the proton beam is absorbed in the
dump and the number of neutrino-producing protons is enhanced relative to
those escaping the source as cosmic rays. For the extreme source of this
type, the observed cosmic rays are all decay products of neutrons
with larger mean-free paths in the dump. The flux for such a source is
shown as the upper horizontal line in Fig.\,6.

\begin{figure}[t!]
\centering\leavevmode
\includegraphics[height=2.5in]{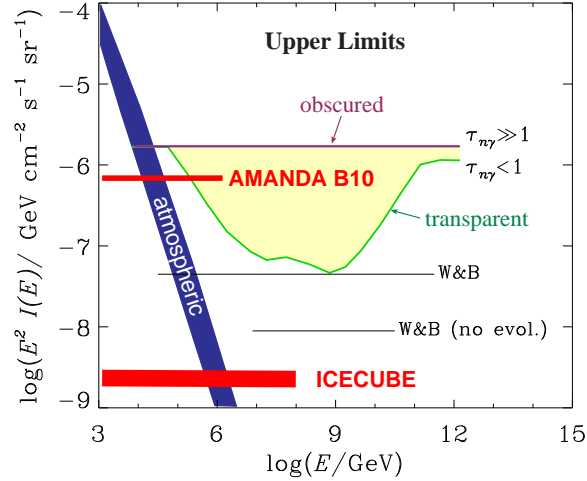}
\caption{The neutrino flux from compact astrophysical accelerators.
Shown
is the range of possible neutrino fluxes associated with the the highest
energy cosmic rays.  The lower line, labeled ``transparent", represents
a source where each cosmic
ray interacts only once before escaping the object.  The upper line,
labeled ``obscured", represents an ideal neutrino source where all
cosmic rays escape in the
form of neutrons.  Also shown is the ability of AMANDA and IceCube to
test
these models.}
\label{six}
\end{figure}

The above limits are derived from the fact that theorized neutrino
sources
do not overproduce cosmic rays. Similarly, observed gamma ray fluxes
constrain potential neutrino sources because for every parent charged
pion
($\pi^{\pm}\rightarrow l^{\pm}+\nu$), a neutral pion and two gamma rays
($\pi^0\rightarrow \gamma + \gamma$) are produced. The
electromagnetic
energy associated with the
decay of neutral pions should not exceed observed astronomical fluxes.
These calculations must take into account cascading of the
electromagnetic
flux in the background photon and magnetic fields. A simple
argument relating high-energy photons and neutrinos produced by secondary
pions can still be derived by relating their total energy and allowing
for
a steeper photon flux as a result of cascading. Identifying the photon
fluxes with those of non-thermal TeV
photons emitted by supernova remnants and blazers, we predict neutrino
fluxes at the same level as the Waxman-Bahcall flux\cite{gammanu}. It is
important to
realize however that there is no evidence
that these are the decay products of $\pi^0$'s. The sources
of the cosmic rays have not been revealed by photon or proton astronomy
\cite{gas1,gas2,gas3,gas4}; see however reference \cite{cangoroo}.

For neutrino detectors to succeed they must be sensitive to the range of
fluxes covered in Fig.\,6. The AMANDA detector has already entered the
region of sensitivity and is eliminating specific models which predict
the largest neutrino fluxes within the range of values allowed by general
arguments. The IceCube detector, now under construction, is sensitive to
the full range of beam dump models,
whether generic as or modeled as active galaxies or gamma ray bursts.
IceCube will reveal the
sources of the cosmic rays or derive an upper limit that will
qualitatively raise the bar for solving the cosmic ray puzzle. The
situation could be nothing but desperate with the escape to top-down
models being cut off by the accumulating evidence that the highest energy
cosmic rays are not photons. In top-down models, decay products
eventually materialize as quarks and gluons that fragment into jets of
neutrinos
and
photons and very few protons.

\section{High Energy Neutrino Telescopes}

Although neutrino telescopes have multiple interdisciplinary science
missions, the search for the sources of the highest-energy cosmic
rays stands out because it clearly identifies the size of the
detector required to do the science\cite{snr1}.

Whereas the science is compelling, the real challenge has been to
develop a reliable, expandable and affordable detector technology.
Suggestions to use a large volume of deep ocean water for high-energy
neutrino
astronomy were made as early as the 1960s. In the case
of the muon neutrino, for instance, the neutrino ($\nu_\mu$)
interacts with a hydrogen or oxygen nucleus in the water and produces
a muon travelling in nearly the same direction as the neutrino. The
blue Cerenkov light emitted along the muon's $\sim$kilometer-long
trajectory is detected by strings of photomultiplier tubes deployed
deep below the surface. With the first observation of neutrinos in
the Lake Baikal and
the (under-ice) South Pole neutrino telescopes, there is optimism
that the technological challenges to build neutrino telescopes can
hopefully be met.

The first generation of neutrino telescopes, launched by the bold
decision of the DUMAND collaboration to construct such an
instrument, are designed to reach a large telescope area
and detection volume for a neutrino threshold of order 10~GeV. The
optical requirements of the detector medium are severe. A large
absorption length is required because it determines the spacings of
the optical sensors
and, to a significant extent, the cost of the detector. A long
scattering length is needed to preserve the geometry of the Cerenkov
pattern. Nature has been kind and offered ice and water as adequate
natural Cerenkov media. Their optical properties are, in fact,
complementary. Water and ice have similar attenuation length, with
the role of scattering and absorption reversed. Optics seems, at
present, to drive the evolution of ice and
water detectors in predictable directions: towards very large
telescope area in ice exploiting the long absorption length, and
towards lower threshold and good muon track reconstruction in water
exploiting the long scattering length.

DUMAND, the pioneering project located off the coast of Hawaii,
demonstrated that muons could be detected by this
technique\cite{dumand}, but the planned detector was never realized. A
detector composed of 96 photomultiplier tubes located deep in Lake
Baikal was the first to demonstrate the detection of neutrino-induced
muons in natural water\cite{baikal2,baikal3}. In the following years,
{\it NT-200} will be operated as a neutrino telescope with an effective
area between $10^3 {\sim} 5\times 10^3$\,m$^2$, depending on energy.
Presumably too small to detect neutrinos from extraterrestrial
sources, {\it NT-200} will serve as the prototype for a larger
telescope. For instance, with 2000 OMs, a threshold of $10 {\sim}
20$\,GeV and an effective area of $5\times10^4 {\sim} 10^5$\,m$^2$, an
expanded Baikal telescope would fill the gap between present
detectors and planned high-threshold detectors of cubic kilometer
size. Its key advantage would be low threshold.

The Baikal experiment represents a proof of concept for deep ocean
projects. These do however have the advantage of larger depth and
optically
superior water. Their challenge is to find reliable and affordable
solutions to a variety of technological challenges for deploying a
deep underwater detector. The European collaborations
ANTARES\cite{antares,antares1,antares2} and
NESTOR\cite{nestor,nestor1,nestor2} plan to deploy
large-area detectors in the Mediterranean Sea within the next year.
The NEMO Collaboration is conducting a site study for a
future kilometer-scale detector in the Mediterranean\cite{NEMO}.

The AMANDA collaboration, situated at the U.S. Amundsen-Scott South Pole
Station, has demonstrated the merits of natural ice as a Cerenkov
detector medium\cite{B4}. In 1996, AMANDA was able to
observe atmospheric neutrino candidates using only 80 eight-inch
photomultiplier tubes\cite{B4}.

\begin{figure}[t]
\centering\leavevmode
\includegraphics[height=3in]{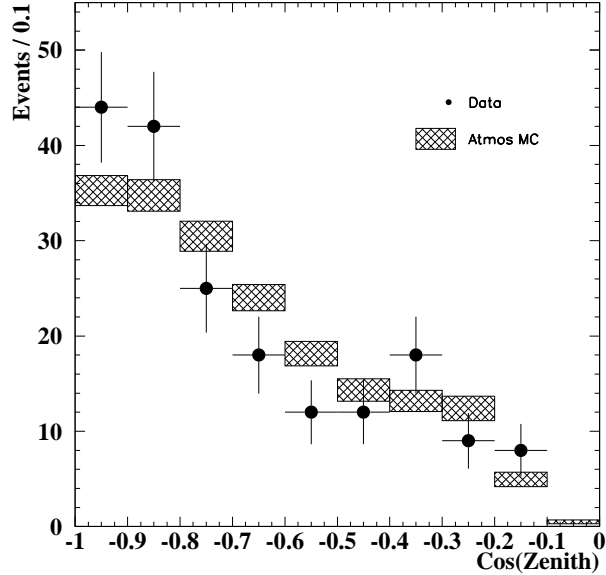}

\caption[]{Reconstructed zenith angle distribution. The
points mark the data and the shaded boxes a simulation of
atmospheric neutrino events, the widths of the boxes
indicating the error bars.
\label{fig:zenith}}
\end{figure}

With 302 optical modules instrumenting approximately 6000 tons of
ice, AMANDA extracted several hundred atmospheric neutrino events
from its first 130 days of data. AMANDA was thus the first
first-generation neutrino
telescope with an effective area in excess of 10,000 square meters
for TeV muons\cite{nature00}.  In rate and all characteristics the
events are
consistent with atmospheric neutrino origin. Their energies are in
the 0.1--1\,TeV range. The shape of the zenith angle distribution is
compared to a simulation of the atmospheric neutrino signal in
Fig.~\ref{fig:zenith}. The variation of the measured rate with zenith
angle is reproduced by the simulation to within the statistical
uncertainty. Note that the tall geometry of the detector strongly
influences the dependence on zenith angle in favor of more vertical
muons.

\begin{figure}[t]
\centering\leavevmode
\includegraphics[height=2.2in]{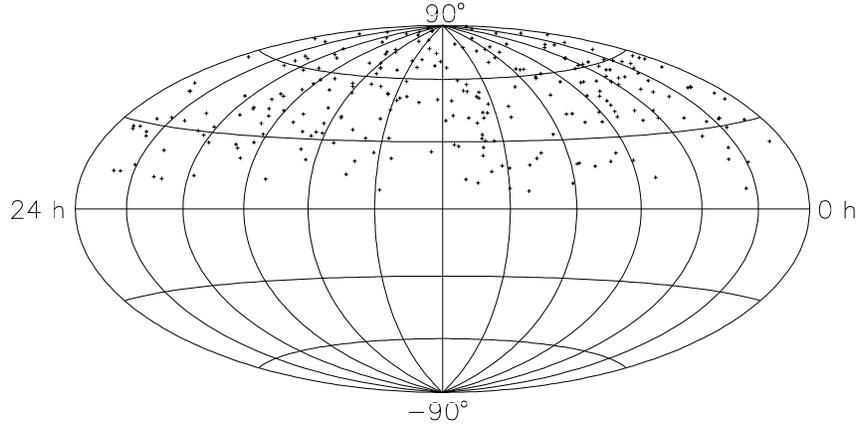}

\caption[]{Distribution in declination and right ascension of the
up-going
events on the sky. \label{fig:skyplot}}
\end{figure}

The arrival directions of the neutrinos are shown in
Fig.~\ref{fig:skyplot}. A statistical analysis indicates no evidence
for point sources in this sample. An estimate of the energies of the
up-going muons (based on simulations of the number of reporting
optical modules) indicates that all events have energies consistent
with an atmospheric neutrino origin. This enables AMANDA to reach a
level of sensitivity to a diffuse flux of high energy
extra-terrestrial neutrinos of order\cite{nature00} $dN/dE_{\nu} =
10^{-6}
E_{\nu}^{-2} \rm\, cm^{-2}\, s^{-1}\,sr^{-1}\, GeV^{-1}, $ assuming
an $E^{-2}$ spectrum. At this level they exclude a variety of
theoretical models which assume the hadronic origin of TeV photons
from active galaxies and blazars\cite{stecker2}. Searches for
neutrinos from gamma-ray bursts, for magnetic monopoles, and for a
cold dark matter signal from the center of the Earth are also in
progress and, with only 138 days of data, yield limits comparable to
or better than those from smaller underground neutrino detectors that
have operated for a much longer period.

In January 2000, AMANDA-II was completed. It consists of 19 strings
with a total of 677 OMs arranged in concentric circles, with the ten
strings from AMANDA forming the central core of the new detector.
First data with the expanded detector indicate an atmospheric
neutrino rate increased by a factor of three, to 4--5 events per day.
AMANDA-II has met the key challenge of neutrino astronomy: it has
developed a reliable, expandable, and affordable technology for
deploying a kilometer-scale neutrino detector named IceCube.

IceCube is an instrument optimised to detect and characterize sub-TeV
to multi-PeV neutrinos of all flavors (see Fig.\,9) from
extraterrestrial sources. It will consist of 80 strings, each with 60
10-inch photomultipliers spaced 17~m apart.
The deepest module is 2.4~km below the surface. The strings are
arranged at the apexes of equilateral triangles 125\,m on a side. The
effective detector volume is about a cubic kilometer, its precise
value depending on the characteristics of the signal. IceCube will
offer great advantages
over AMANDA II beyond its larger size: it will have a much higher
efficiency to reconstruct tracks, map showers from electron- and
tau-neutrinos (events where both the production and decay of a $\tau$
produced by a $\nu_{\tau}$ can be identified) and, most
importantly, measure neutrino energy. Simulations indicate that the
direction of muons can be determined with sub-degree accuracy and
their energy measured to better than 30\% in the logarithm of the
energy. Even the direction of showers can be reconstructed to better
than 10$^\circ$ in both $\theta$, $\phi$ above 10\,TeV. Simulations
predict a linear response in energy of better than 20\%. This has to
be contrasted with the logarithmic energy resolution of
first-generation detectors. Energy resolution is critical because,
once one establishes that the energy exceeds 100~TeV, there is no
atmospheric neutrino background in a kilometer-square detector.

\begin{figure}[t!]
\centering\leavevmode
\includegraphics[height=1.8in]{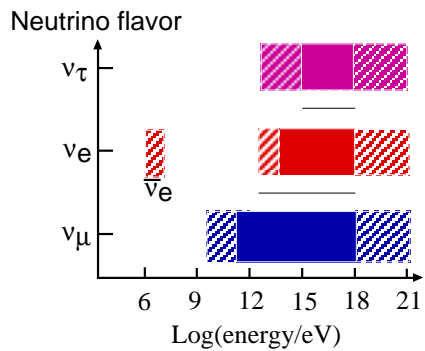}
\caption{Although IceCube detects neutrinos of any flavor above a
threshold of $\sim 0.1$\,TeV, it can identify their flavor and measure
their energy in the ranges shown.}
\end{figure}

At this point in time, several of the new instruments, such as the
partially deployed Auger array and HiRes to Magic to Milagro and
AMANDA~II, are less than one year from delivering results. With rapidly
growing observational capabilities, one can express the realistic hope
that the cosmic ray puzzle will be solved soon. The solution will almost
certainly reveal unexpected astrophysics, if not particle physics.

For a recent review of neutrino astronomy and its relationship to cosmic rays, see 
Ref.~\cite{review}.

\section*{Acknowledgments}
This work was supported in part by DOE grant No. DE-FG02-95ER40896 and
in part by the
Wisconsin Alumni Research Foundation.

\end{document}